\documentstyle[twocolumn,aps]{revtex}

\begin{document}

\draft
\preprint{\begin{tabular}{l} 9612dd\end{tabular}}

\title{Topological Invariants, Instantons and Chiral Anomaly on
Spaces with Torsion} 

\author{Osvaldo Chand\'{\i}a$^{a,b}$ and Jorge Zanelli$^{a,c}$}

\address{$^{(a)}$ Centro de Estudios Cient\'{\i}ficos de Santiago,
Casilla 16443, Santiago, Chile\\ 
$^{(b)}$ Departamento de F\'{\i}sica, Facultad de Ciencias,
Universidad de Chile, Casilla 653, Santiago, Chile\\
$^{(c)}$ Departamento de F\'{\i}sica, Universidad
de Santiago de Chile, Casilla 307, Santiago 2, Chile.}
\maketitle

\begin{abstract}
In a spacetime with nonvanishing torsion there can occur
topologically stable configurations associated with the frame bundle
which are independent of the curvature. The relevant topological
invariants are integrals of local scalar densities first discussed by
Nieh and Yan (N-Y). In four dimensions, the N-Y form $N=(T^a \mbox{\tiny $\wedge$} T_a - R_{ab}\mbox{\tiny $\wedge$} e^a\mbox{\tiny $\wedge$} e^b)$ is the only closed 4-form invariant under local Lorentz
rotations associated with the torsion of the manifold. The integral
of $N$ over a compact D-dimensional (Euclidean) manifold is shown to
be a topological invariant related to the Pontryagin classes of
$SO(D+1)$ and $SO(D)$. An explicit example of a topologically
nontrivial configuration carrying nonvanishing instanton number
proportional to $\int N$ is costructed.  The chiral anomaly in a
four-dimensional spacetime with torsion is also shown to contain a
contribution proportional to $N$, besides the usual Pontryagin
density related to the spacetime curvature. The violation of chiral
symmetry can thus depend on the instanton number of the tangent frame
bundle of the manifold. Similar invariants can be constructed in
$D>4$ dimensions and the existence of the corresponding nontrivial
excitations is also discussed.
\end{abstract}

\section{Introduction}
In the traditional approach to gravitation theory, torsion plays no
significant role in the spacetime geometry. Torsion is commonly set
equal to zero from the start and there seems to be no compelling
experimental reason to relax this condition. In a more geometric
approach, however, the affine and metric properties of the  spacetime
geometry are independent notions and should therefore be described by
dynamically independent fields: the spin connection, ${\omega^a}_b$,
and the local frames (vielbein), $e^a$, respectively \cite{zum}.  In
the
tradition of General Relativity these two fields are assumed to be
linked by the torsion-free condition $T^a =0$, where the torsion
2-form is defined by
\begin{equation}
T^a = d e^a + {\omega^a}_b \mbox{\tiny $\wedge$} e^b.
\label{torsion}
\end{equation}

This expression is similar to that of the curvature 2-form,
\begin{equation}
{R^a}_b = {d \omega^a}_b +{\omega^a}_c \mbox{\tiny $\wedge$} {\omega^c}_b,
\label{curvature}
\end{equation}
whose vanishing is not to be imposed a priori. 

From these two expressions, curvature seems to be more fundamental
than torsion: the definition (\ref{curvature}) depends on the
existence of the connection field alone, whereas torsion depends on
both the connection and the vielbein.  On the other hand, since on
any smooth metric manifold a local frame (vielbein) is necessarily
always defined, torsion can exist even if the connection vanishes.
This implies that in a geometric theory of spacetime the local frame
structure is as basic a notion as the connection and, therefore,
torsion and curvature should be treated on a similar footing. 

From a group theoretic point of view, the curvature 2-form is the
commutator of the covariant derivative for the connection of the
group of rotations on the tangent space of the manifold ($SO(D)$ or
$SO(D-1,1)$, for Euclidean or Minkowskian signature, respectively.\footnote{Here we will assume the signature to be
Euclidean. Whenever spacetime is mentioned, the appropriate Wick
rotation will be assumed.}) This is reflected by the fact that the curvature depends on the group connection ${\omega^a}_b$ alone. In contrast, no analogous simple geometric interpretation can be assigned to torsion. [For an discussion on this point, see section II, below.] This is perhaps one reason why torsion has been perceived as less fundamental than curvature since the early days of General Relativity \cite{C-E}. Nevertheless, torsion appears rather naturally in the commutator of two covariant derivatives for the group of diffeomorphisms of a manifold in a coordinate basis \cite{lovelock},

\begin{equation}
[\nabla_{\mu}, \nabla_{\nu}]V^A = -T^\lambda_{\mu \nu}
\nabla_{\lambda}V^A + R^A_{B \mu \nu} V^B,
\label{commutator}
\end{equation}
where $V^A$ represents any tensor (or spinor) under diffeomorphisms
or under the group of tangent rotations, and $R^A_B$ is the curvature
tensor in the corresponding representation. Here curvature and torsion play quite different roles: $T^\lambda_{\mu \nu}$ is the structure function for the diffeomorphism group and $R^A_{B \mu \nu}$ is a central charge. From 
this expression it is clear that one can consider equally well spaces
with curvature and no torsion, and ``teleparallelizable" spaces with
zero curvature and nonvanishing torsion. Both possibilities are
special cases of the generic situation.

Another realm where curvature plays an important role is in the
characterization of the topological struture of the manifold. It is a
remarkable result of differential geometry that certain global
features of a manifold are determined by some local functionals of
its intrinsic geometry. The four-dimensional Pontryagin and Euler
classes,
\begin{equation}
P_4 = \frac{1}{8\pi^2}\int_{M_4}R^{ab}{\mbox{\tiny $\wedge$}}R_{ab},
\label{p4}
\end{equation}

\begin{equation}
E_4 = \frac{1}{32{\pi}^2} \int_{M_4}\epsilon_{abdc} R^{ab}\mbox{\tiny
$\wedge$}R^{cd},
\label{e4}
\end{equation}
are well known examples. For compact manifolds in four dimensions
$P_4$ and $E_4$ take integer values that label topologically distinct
four-geometries. Although these topological invariants are given in
terms of local functions, their values depend on the global
properties of the manifold. These topological invariants are expected
to be related to physical observables as for instance in the case of
anomalies. The Pontryagin class can be defined for any compact gauge
group $G$, on any even-dimensional compact manifold,
\begin{equation}
P_{2n}[G]= \frac{1}{2^{n+1}\pi^n}\int_{M_{2n}}
\{\underbrace{F\mbox{\tiny $\wedge$} \cdots \mbox{\tiny
$\wedge$}F}_{n} \},
\label{G}
\end{equation}
where $F$ is the curvature 2-form for the group $G$ whose generators
are normalized so that $Tr \{G_a G_b \} = \delta_{ab}$, and the
braces $\{...\}$ indicate a particular product of traces of products
of  $F$'s (see \cite{nakahara}). Since the curvature 2-form for the
manifold ($R^{ab}$) in the standard representation is antisymmetric,
the Pontryagin form {\em of the manifold}, $P_{D}[SO(D)]$ is only
defined for $D=4n$. In contrast with the Pontryagin forms, the Euler
form cannot be defined for a generic gauge group $G$.

Invariants analogous to these, constructed using the torsion tensor
are less known. The lowest dimensional torsional invariant is the
4-form first discussed by Nieh and Yan (N-Y)\cite{NY}, 
\begin{equation}
N=T^a \mbox{\tiny $\wedge$} T_a - R_{ab}\mbox{\tiny $\wedge$}
e^a\mbox{\tiny $\wedge$} e^b.
\label{NY}
\end{equation}
This is the only nontrivial locally exact 4-form which vanishes in
the absence of torsion and is clearly independent of the
Pontryagin and Euler densities. In any local patch where the vierbein
is well defined, $N$ can be written as
\begin{equation}
N= d(e^a \mbox{\tiny $\wedge$}T_a),
\label{eT}
\end{equation}
and is therefore locally exact. More explicitly, if $N$ and $ N'$ are
the N-Y densities for ($\omega$, $e$), and $(\omega', e')$, where
$\omega' = \omega + \lambda$, $e' =e + \zeta $, then $\Delta= N - N'$
is locally exact (a total derivative). If the deformation between
$\omega$ and $\omega '$ is globally continuous,
$\Delta$ is globally exact. Therefore $\int N$ is a topological
invariant quantity in the same sense as the Pontryagin and Euler
numbers. Similar invariants can be defined in higher dimensions as
discussed in \cite{MZ}.

The 3-form $e^a\mbox{\tiny $\wedge$} T_a$ is a Chern-Simons-like form that can be used
as a Lagrangian for the dreibein in three dimensions. The dual of
this 3-form in four dimensions is also known as the totally
antisymmetric part of the torsion (contorsion) and is sometimes also
referred to as H-Torsion,
\begin{equation}
e^a\mbox{\tiny $\wedge$} T_a \sim \epsilon^{\mu \nu \lambda \rho} T_{\nu \lambda \rho}.
\end{equation}
This component of the torsion tensor is the one that couples to the
spin 1/2 fields \cite{obukhov}. This is one of the irreducible pieces
of the first Bianchi identity. In a metric-affine space, the 1st
Bianchi identity can be decomposed according to 16 = 9 + 6 + 1. And
the `1' is that corresponding to (\ref{eT}) \cite{Hehl}. 

In the next section $N$ is shown to be related to the Pontryagin
class, sheding some light on the origin of its topological nature.
Section III contains the construction of a field configuration that
exhibits the relevant instanton number. In section IV, the
contribution of $N$ to the chiral anomaly and the corresponding index
theorem are discussed. A general discussion, in particular, about the
possibility of having similar invariants in higher dimensions, are contained in section V.

\section{Relation to the Pontryagin class}

It seems natural to investigate the extent to which the Nieh-Yan
invariant (\ref{NY}) is analogous to the Pontryagin and Euler
invariants. In particular, it would be intersting to know whether the
integral of $N$ over a compact manifold has a discrete spectrum as is
the case for $E_4$ and $P_4$. 

This question can be answered by embedding the group of rotations on
the tangent space, $SO(4)$ into $SO(5)$. This can be done quite
naturally combining the spin connection and the vierbein together in
a connection for $SO(5)$ in the form \cite{MZ,BTZ,G}
\begin{equation}
W^{AB} = \left[
\begin{array}{cc}
\omega^{ab} & \,\,\frac{1}{l}e^a\\
-\frac{1}{l}e^b & \,\,0
\end{array} \right],
\label{embedding}
\end{equation}
where $a, b =1, 2,\cdots 4$ $A, B =1, 2,\cdots 5$. Note that the
constant $l$ with dimensions of length has been introduced to match
the standard units of the connection ($l^{-1}$) and those of the
vierbein ($l^0$). In the usual embedding of the Lorentz group into
the (anti-) de Sitter group, $l$ is called the radius of the universe
and is related to the cosmological constant ($|\Lambda|= l^{-2}$).

The curvature 2-form constructed from $W^{AB}$ is
\begin{eqnarray}
F^{AB} & = & dW^{AB} +W^{AC}\mbox{\tiny $\wedge$} W^{CB}\nonumber \\
	   & = & \left[ \begin{array}{cc}
		R^{ab} - \frac{1}{l^2}e^a \mbox{\tiny $\wedge$} e^b &
\frac{1}{l}T^a \\
			-\frac{1}{l}T^b				& 0
		\end{array} \right].
\label{F}
\end{eqnarray}

It is then direct to check that the Pontryagin density for $SO(5)$ is
the sum of the Pontryagin density for $SO(4)$ and the Nieh-Yan
density,

\begin{equation}
F^{AB} \mbox{\tiny $\wedge$} F_{AB} = R^{ab}\mbox{\tiny $\wedge$}
R_{ab} + \frac{2}{l^2}[T^a\mbox{\tiny $\wedge$} T_a -
R^{ab}\mbox{\tiny $\wedge$} e_a \mbox{\tiny $\wedge$} e_b].
\end{equation}
This shows, in particular, that 
\begin{equation}
\frac{2}{l^2}\int_{M_4} N= P_4[SO(5)] - P_4[SO(4)],
\label{NPP}
\end{equation}
is indeed a topological invariant, as it is the difference of two
Pontryagin classes. 

From (\ref{NPP}) one can directly read off the spectrum of $\int N$.
As is well known, the Pontryagin class of $P_{2n}[G]$ takes on
integer values (the instanton number) of the corresponding homoptopy
group, $\Pi_{2n-1}(G)$ (see, e.g., \cite{nakahara}). In the case at
hand, $\Pi_3(SO(5)) =$ {\bf Z} and  $\Pi_3(SO(4)) =$ {\bf Z + Z}.
Thus, the integral of the Nieh-Yan invariant over a compact manifold
$M$ must be a function of three integers,
\begin{equation}
\int_M N = \mbox{const.}\times (z_1 + z_2 + z_3), \;\;\; z_i \in
\mbox{{\bf Z}}.
\label{spectrum}
\end{equation}

\section{Instanton}

It is of interest to construct an example geometry with nonvanishing
$\int_M N$. As it is seen from (\ref{eT}), the integral (\ref{spectrum}) can be evaluated integrating of the 3-form $e_a\mbox{\tiny $\wedge$}T^a$ over the boundary $\partial M$. 

A particular example of a geometry characterized by nonvanishing
$\int N$ is easily constructed using the fact that $N$ may be nonzero
even if the curvature vanishes. The simplest example occurs in
I$\!$R$^4$, where the connection can be chosen to vanish everywhere
$\omega^{ab} =0$. Consider now a vierbein field that approaches a
regular configuration as $r \rightarrow \infty$. The question is how
to cover the sphere at infinity ($S^3_{\infty}$) with an everywhere
regular set of independent vectors. 

It is a classical result on fibre bundles that $S^3$ is
parallelizable, i.e., there exist three linearly independent globally
defined vector fields over the sphere \cite{adams}. Using this fact
it is possible to take one of the vierbein field along the radius
($e^r $) and the other three tangent to the $S^3$. Defining the
sphere through its embedding in I$\!$R$^4$, $x^2 + y^2 + z^2 +u^2
=r^2$, we chose on its surface

\begin{eqnarray}
e^r &=& \frac{l}{r}dr \nonumber \\
e^1 &=&  \frac{l}{r^2}( ydx - xdy - udz + zdu) \nonumber \\
e^2 &=&  \frac{l}{r^2}(-zdx - udy + xdz + ydu) \label{dreibein}\\
e^3 &=&  \frac{l}{r^2}( udx - zdy + ydz - xdu). \nonumber 
\end{eqnarray}

These fields are well defined for $r\neq 0$ and can be smoothly
continued inside the sphere, for instance rescaling it by a function
that vanishes as $r\rightarrow 0$ and  approaches 1 for $r\rightarrow
\infty$. In any case, it is clearly impossible to do this without
producing a singular point where $e^a$ vanishes.  The integral of
$e_a{\mbox{\tiny $\wedge$}}T^a$ over a sphere of any radius is

\begin{equation}
\frac{1}{l^2}\int_{S^3} e_a {\mbox{\tiny $\wedge$}} de^a = 3\cdot 2\cdot 2\pi^2.
\label{int}
\end{equation}

Thus, using (\ref{eT}), one concludes that the above result
corresponds to the integral of the Nieh-Yan form over I$\!$R$^4$. The
factor 3 comes from the fact that there are 3 independent fields
summed in the integrand of (\ref{int}). Each term in the sum
contributes twice the area of the unit 3-sphere ($2\pi^2$). 

Configurations with other instanton numbers can be easily generated
by simply choosing different winding numbers for each of the three
tangent vectors $e^i$. In the example above each of these vectors
makes a complete turn around the equatorial lines defined by the
planes $x=y=0$, $x=z=0$, and $x=u=0$, respectively. We are thus led
to conclude that, in general,

\begin{equation}
\frac{1}{l^2} \int_M N = 4\pi^2 (z_1 + z_2 + z_3),\;\;\; z_i \in
\mbox{{\bf Z}}.
\label{integral}
\end{equation}

The instanton presented here is analogous to the one discussed by
D'Auria-Regge \cite{DR}. Theirs is also associated to a singularity
in the vierbein structure of the manifold, but has vanishing N-Y
number and nonzero Pontryagin and Euler numbers.

\section{Chiral Anomaly}

It is well known that the existence of anomalies can be atributed to
the topological properties the background where the quantum system is
defined. In particular, for a masless spin one-half field in an
external (not necessarily quantized) gauge field $G$, the anomaly for
the conservation law of the chiral current is proportional to the
Pontryagin form for the gauge group,
\begin{equation}
{\partial}_{\mu}<J_5^{\mu}> = \frac{1}{4{\pi^2}} Tr F \mbox{\tiny $\wedge$} F.
\label{anomaly}
\end{equation}

The question then naturally arises as to whether the torsional
invariants can produce similar physically observable effects
\cite{MZ}. 

Kimura \cite{kimura}, Delbourgo and Salam \cite{DS}, and Eguchi and
Freund \cite{EF} evaluated the quantum violation of the chiral
current
conservation in a four dimensional Riemannian background {\em without
torsion}, finding it proportional to the Pontryagin density of the
manifold,

\begin{equation}
{\partial}_{\mu}<J_5^{\mu}> =  \frac{1}{8{\pi}^2}R^{ab}{\mbox{\tiny
$\wedge$}} R_{ab}.
\label{grav}
\end{equation}

This result was also supported by the computation of Alvarez-Gaum\'e
and Witten \cite{AW}, of all possible gravitational anomalies and the
Atiyah-Singer index for the Dirac operator for massless fermions in a
curved background, and the complete study of consistent nonabelian
anomalies on arbitrary manifolds by Bonora, Pasti and Tonin
\cite{BPT}.

It has been sometimes argued that the presence of torsion could not
affect
the chiral anomaly (see, e.g., \cite{BPT,WZ,KY,mavromatos}). This is
motivated by the fact that the Pontryagin number is insensitive to
the presence of torsion, as it is obvious from
Eqs.(\ref{curvature},\ref{p4}). This does not prove, however, that
the anomaly is given by the
Pontryagin class and nothing else. 

As for the Atiyah-Singer index theorem, it is fairly clear that the
difference between the number of left- and right-handed zero modes
should not jump under any continuous deformation of the geometry.
Therefore the index could not change under adiabatic inclusion of
torsion in the connection. However, nothing can be said {\it a
priori} about the changes of the index under {\em discontinuous}
modifications in the torsion, as it might happen if flat spacetime is
replaced by one containing a topologically nontrivial configuration. 

The integral of an anomaly must be a topological invariant
\cite{zumino} and therefore the assertion above would be true if
there were no other independent topological invariants that could be
constructed out of the torsion tensor. 

Direct computations of the chiral anomaly in spaces with torsion were
first done by Obukhov \cite{obukhov}, and later by Yajima and
collaborators \cite{KY,Y}. These authors find a number of
torsion-dependent contributions to the anomaly which are not clearly
interpreted as densities of topological invariants.  In a related
work, Mavromatos \cite{mavromatos} calculates the Atiyah-Singer index
of the Dirac operator in the presence of curl-free H-torsion. He
finds a contribution which is an exact form by virtue of his
assumption (curl-free H-torsion amounts to assuming $d(e^a \mbox{\tiny $\wedge$} T_a)
=0$) and is therefore dropped out. 

In all the previous cases \cite{obukhov,KY,Y} (and in
\cite{mavromatos} if one doesn't assume $dH=0$), the N-Y term appears
among many other.  Many of these torsional pieces, including the N-Y
term, are divergent when the regulator is removed, which was
interpreted as an indication that these terms were a regulator
artifact and should therefore be ignored. 

Here we recalculate the anomaly with the Fujikawa method
\cite{fujikawa,D-R} and explicitly show the dependence of the anomaly
on the N-Y 4-form.

Consider a massless Dirac spinor on a curved background with
torsion. The action is
\begin{equation}
S=\frac{i}{2} \int d^4x e {\bar{\psi}}{\not \!
\nabla}\psi
+ h.c.,
\label{action}
\end{equation}
where the Dirac operator is given by
\begin{equation}
{\not \! \nabla} = {e_a}^{\mu} {\gamma}^a \nabla_{\mu},
\end{equation} 
here ${e_a}^{\mu}$ is the inverse of the tetrad ${e^a}_{\mu}$,
${\gamma}^a$ are the Dirac gamma matrices and $\nabla_\mu$ is the
covariant derivative for the $SO(4)$-connection in the appropriate
representation. 

This action is invariant under rigid chiral transformations
\begin{equation}
\psi {\longrightarrow} e^{i\varepsilon {\gamma}_5}\psi
\label{chiral}
\end{equation}
where $\varepsilon$ is a real constant parameter. This symmetry leads
to the classical conservation law 
\begin{equation}
{\partial}_{\mu}J^{\mu}_5 = 0
\label{conservation}
\end{equation}
where $J^{\mu}_5 = e {e_a}^{\mu} {\bar{\psi}} {\gamma}^a
{\gamma}_5 \psi$.

The chiral anomaly is given by  
\begin{equation}
{\partial}_{\mu} <J^{\mu}_5> = {\cal A}(x),
\label{naive}
\end{equation}
where 
\begin{equation}
{\cal A}(x) = 2\sum_{n} e(x) {\psi}_n^{\dagger} {\gamma}_5
{\psi}_n.
\label{infinite}
\end{equation}
With the standard regularization $\cal A$ is
\begin{equation}
{\cal A}(x) = 2\lim_{y\rightarrow x} \lim_{M\rightarrow
{\infty}} Tr \left[{\gamma}_5 \exp(\frac{{\not \!
\nabla}^2}{M^2})\right] {\delta}(x,y).
\label{reg}
\end{equation}

The square of the Dirac operator is given by
\begin{equation}
{\not \! \nabla}^2 = \nabla^{\mu}\nabla_{\mu} - e^{\mu}_a e^{\nu}_b
e^{\lambda}_c J^{ab} T^c_{\mu \nu}\nabla_{\lambda} + 
\frac{1}{2} e^{\mu}_a e^{\nu}_b J^{ab}J^{cd} R_{cd \mu \nu},
\end{equation}
where ${J}_{ab}=\frac{1}{4} [{\gamma}_a , {\gamma}_b ]$ is the
generator of $SO(4)$ in the spinorial representation.

The Dirac delta on a curved background is represented by
\begin{equation}
\delta (x,y) = \int \frac{d^4k}{(2\pi )^4} e^
{k^{\mu}{\nabla}_{\mu}{\Sigma}(x,y)},
\label{28}
\end{equation}
where ${\Sigma}(x,y)$ is the geodesic biscalar \cite{dewitt}. Applying the operator $\exp(\frac{{\not \! {\nabla}}^2}{M^2})$ on (\ref{28}), taking the limit $y{\rightarrow}x$, and tracing over spinor indices, one finds
\begin{eqnarray}
{\cal A} & = &\frac{1}{8{\pi}^2}[R^{ab}{\mbox{\tiny $\wedge$}}R_{ab} +
2M^2(T_a \mbox{\tiny $\wedge$} T^a - R_{ab}\mbox{\tiny $\wedge$} e^a\mbox{\tiny $\wedge$} e^b] \nonumber \\
 & & + O(M^{-2}).
\label{result}
\end{eqnarray}

The leading contribution of torsion to the anomaly, $\frac{M^2}{ \pi^2}N$, diverges as the regulator is removed in agreement with the results of \cite{obukhov,KY,Y,mavromatos} \footnote{In a Pauli-Villars regularization scheme, such divergent terms would be eliminated by an appropriate choice of regulator mass parameters. This scheme, however, rests on the assumption that at high energy spacetime has Poincar\'e invariance, but this is not a trivial assumption in the presence of gravity. We thank G. 't Hooft for pointing this out to us.}. 

A finite result would be obtained if the vierbein were rescaled as
\begin{equation}
e^a{\longrightarrow} \tilde{e}^a = \frac{1}{M} e^a.
\end{equation}
In that case the expression for the anomaly in the limit $M\rightarrow \infty$ becomes
\begin{equation}
{\cal A}(x) = {\frac{1}{8\pi^2}}\left[
R^{ab}{\mbox{\tiny $\wedge$}}R_{ab} + 2
(T^a{\mbox{\tiny $\wedge$}}T_a - R_{ab}{\mbox{\tiny
$\wedge$}}e^a{\mbox{\tiny $\wedge$}}e^b)\right].
\label{result'}
\end{equation}

It is interesting to observe that if in the earlier results of refs. (\cite{obukhov,KY,Y}), one makes the same rescaling, all but one of
the torsional contributions to the anomaly vanish in the limit
$M\rightarrow \infty$. The remaining term is $N$.

\section{Discusion}

\subsection{Higher dimensions}

Topological invariants associated to the spacetime torsion exist in
higher dimensions whose occurrence, however is very hard to predict
for arbitrary $D$ \cite{MZ}. An obvious family of these invariants for $D=4k$, is of the form $N^k= N\mbox{\tiny $\wedge$} N\mbox{\tiny $\wedge$} \cdots N$, but there are others
which do not fall into this class. For example, in 14 dimensions,
the 14-form $(T_a\mbox{\tiny $\wedge$} {R^a}_b\mbox{\tiny $\wedge$} {R^b}_c\mbox{\tiny $\wedge$} {R^c}_d\mbox{\tiny $\wedge$} e^d)\mbox{\tiny $\wedge$} (T_a\mbox{\tiny $\wedge$}
{R^a}_b\mbox{\tiny $\wedge$} e^b)$ is a locally exact. The number $n(D)$ of independet torsional invariants for a given dimension is as follows:  
\\

\begin{tabular}{|l|l|l|l|l|l|l|l|} \hline
$D$    &2  &4 &6 &8 &10 &12 &14 \\ \hline
$n(D)$ &0  &1 &0 &4 &0  &12 &1  \\ \hline
\end{tabular}
\\

The instanton constructed here is easily generalized for $D=8$, where there are four N-Y forms (the wedge product is implicitly assumed),
\begin{eqnarray} 
N_1 & = &  N^2  \nonumber \\
N_2 & = & ({R^a}_b {R^b}_a)N \nonumber \\
N_3 & = & 4(T_a {R^a}_b e^b)(T_a e^a) + (T_a T^a)^2 - (e_a {R^a}_b
e^b)^2 \nonumber \\
N_4 & = & T_a {R^a}_b {R^b}_c T^c - e_a {R^a}_b {R^b}_c {R^c}_d e^d.
\label{D=8}
\end{eqnarray}

Of all these, only $(T_a T^a)^2 = d[e^a T_a T^b T_b]$ survives if the space is assumed to be curvature-free. The integration over a seven sphere $x_1^2 + \cdots + x_8^2 =r^2$ embedded on I$\!$R$^8$ can be easily performed using a frame formed by one radial 1-form ($e^r$) and seven orthonormal fields ($e^i$), tangent to $S^7$. The $e^i$'s are generated using the canonical isomorphism between I$\!$R$^8$ and the octonion algebra: multiplying $e^r$ by each of the seven generators of the algebra, and seven orthonormal fields tangent to the sphere are produced. The first one is
\begin{eqnarray}
e^1 & = & -x_2 dx_1 + x_1 dx_2 - x_4 dx^3 + x_3 dx^4 \nonumber \\
    &   & -x_6 dx^5 + x_5 dx^6 - x_8 dx^7 + x_7 dx^8,
\label{S7}
\end{eqnarray}
and the rest are similarly obtained. The integral is thus a combinatorial factor times the volume of the $S^7$     ($\frac{\pi^4}{3}$).

In 8 dimensions the integral of $N^2$ is not simply equal to the
difference of the Pontryagin classes of $SO(9)$ and $SO(8)$, as one
could naively expect by analogy with the case $D=4$. The Pontryagin
density of $SO(9)$ is
\begin{eqnarray}
Tr(F^4) - \frac{1}{2} (Tr(F^2))^2 & = & Tr(R^4) - \frac{1}{2}
(Tr(R^2))^2  \nonumber \\ 
+ \frac{4}{l^4}[(2T_a {R^a}_b e^b)(T_a e^a)& + &(e_a {R^a}_b e^b)N]
\nonumber \\ 
+ \frac{4}{l^2}[ e_a {R^a}_b {R^b}_c {R^c}_d e^d & + & \frac{1}{2}
Tr(R^2)N - T_a {R^a}_b {R^b}_c T^c].
\label{so(9)}
\end{eqnarray}
The first two terms are the Pontryagin form of $SO(8)$, but the terms
that depend on the torsion vanish for ${R^a}_b =0$. 

Note that a construction  similar to the one for the 3- and 7-spheres
cannot be repeated in any other dimension because only the 1- 3- and
7- spheres admit a globally defined basis of tangent vector fields \cite{adams}. In general, the maximum number of independent global vectors that can be defined on $S^{n-1}$ is given by Radon's formula
\cite{husemoller},
\begin{equation}
\rho_n = 2^c +8d -1,
\label{rho}
\end{equation} 
where $n$ is written as
\begin{equation}
n = (\mbox{odd integer}) 2^c 16^d,
\label{}
\end{equation}
with $c\leq 3$ and $d$ positive integers. From this formula, it is
clear that for all odd-dimensional spheres $\rho_n \geq 1$, while for
even-dimensional spheres (odd $n$), $\rho_n =0$. 

Thus, only it is only in four dimensions that the N-Y class can be computed in a curvature-free background.

\subsection{Anomaly}

In Section IV we argued that the anomaly could be made finite if one were to rescale the tetrad as $e^a_{\mu}\rightarrow \tilde{e}^a_{\mu} = M e^a_{\mu}$. Two remarks are in order: First, it should be stressed that this is the only rescaling that is needed to yield a finite result. Second, the Lagrangian for the tetrad field has not been discussed and therefore the replacement $e\rightarrow \tilde{e}$ is purely formal and can have no physical consequences as long as its dynamics is not specified. 

In our analysis $e$ is an external (classical) background field. One
could view the rescaling of the vierbein as an invariance of the
action, provided the Dirac field is suitably rescaled as well. This
transformation was also considered by Nieh and Yan in \cite{NY'}.
However, ir order for this invariance of the action to be interpreted
as a symmetry generated by charges acting on the fields, one should
include a scale invariant Lagrangian for $e$. 

The rescaled vierbein $\tilde{e}$ has units of mass and is therefore
of the same canonical dimension as the connection. If $\tilde{e}$ is
to be regarded as part of a connection of $SO(5)$, the limit
$M\rightarrow \infty$ keeping $Ml$ fixed could be interpreted as a
way to turn the $SO(4)$ invariant action (\ref{action}) into that for
a spinor minimally coupled to a $SO(5)$ connection. In this case, the
chiral anomaly is then given by $P_4[SO(5)]$, which is precisely
(\ref{result'}).

\subsection{Index}

The Atiyah-Singer index for the Dirac operator in the absence of
torsion is given by the Pontryagin number. Obviously, as $P$ is
independent of the vierbein, its invariance  under continuous
deformations of the geometry also allows for continuous deformations
of the local frames and, in particular, for the addition of torsion.
A different issue is whether the presence of torsion can affect the
index of the Dirac operator through these invariants. 

In \cite{mavromatos} it is shown that there is a torsional
contribution to the index although it is set equal to zero by the aditional requirement of curl-free H- torsion, and our result (\ref{result'}) agrees with that conclusion. The expression of
the anomaly (\ref{result'}) indicates that if the index is calculated for a $SO(5)$ connection, the result would reproduce our expression \cite{cz}.

Note added: In the process of writing this article, we received a
draft by Y. Obukhov, E. Mielke, J. Budczies and F. W. Hehl
\cite{OMBH} where the instanton of Section III is reobtained in a somewhat different analysis, and our result for the anomaly (\ref{result}) is also found in the heat kernel approach.

\noindent {\large {\bf Acknowledgments}}\\

We are deeply indebted to R. Baeza, L. Bonora,  M. Henneaux, G. 't
Hooft and F. Urrutia for patiently following our arguments and for
their helpful comments. We would also like to thank J. Alfaro, M.
Ba\~nados, N. Bralic,  J. Gamboa, A. Gomberoff, F. M\'endez and R.
Troncoso for useful comments and criticism. We gratefully acknowledge helpful discussions and correspondence with F. W. Hehl and Y. Obukhov, and for letting us have a copy of their unpublished
draft \cite{OMBH}. This work was partially supported by grants
1960229, 1940203 and 1970151 from FONDECYT-Chile, and grant 27-9531ZI
from DICYT-USACH. The institutional support by a group of Chilean
companies (EMPRESAS CMPC, BUSINESS DESIGN ASS., CGE, CODELCO, COPEC,
MINERA ESCONDIDA, NOVAGAS and XEROX-Chile) is also recognized.


\end{document}